\documentclass[preprint,aps,prb,12pt,showpacs,floatfix]{revtex4}
\usepackage{epsfig}
\begin{document}
\input epsf

\title{
Local order in aqueous solutions of rare gases and the role of 
the solute concentration: a computer simulation 
study with a polarizable potential.}
\author{Paola~Cristofori}
\author{Paola~Gallo}\author{Mauro~Rovere}\email[Author to whom 
correspondence should be addressed: ]{rovere@fis.uniroma3.it}
\affiliation{Dipartimento di Fisica, 
Universit\`a ``Roma Tre'', \\ INFM Roma Tre and 
Democritos National Simulation Center \\
Via della Vasca Navale 84, 00146 Roma, Italy.}

\begin{abstract}

Aqueous solutions of rare gases are studied by computer simulation 
employing a polarizable potential for both water and solutes.
The use of a polarizable potential allows to study 
the systems from ambient to supercritical conditions for water.
In particular the effects of increasing the concentration and the size
of the apolar solutes are considered in an extended range of 
temperatures. 
By comparing the results at increasing temperature
it appears clearly the change of behaviour from the tendency to demix
at ambient conditions to a regime of complete solubility
in the supercritical region. In this respect
the role of the hydrogen bond network of water is evidenced.

\end{abstract}


\maketitle
\date{\today}

\section{Introduction}
The study of apolar aqueous solutions is of relevant interest 
for understanding basic effects connected to the properties 
of water as a solvent. In presence of an apolar
solute reorganization of water solvent is observed around
the hydrophobic solute molecules. The ordering of water
causes a decrease of entropy in competition with 
the enthalpic term which favours the solvation.     
The apolar molecules of solute tend to aggregate to
reduce the local order of the water molecules. 
The balance between the entropic and the enthalpic
terms determines the hydrophobic 
hydration phenomena~\cite{eley,frank,ben-naim}. 
The hydrophobic effect
plays an important role in chemistry, biology,
chemical engineering. 
In this respect
aqueous solutions of rare gases
are considered both in experiments and in computer simulations
as prototype models 
for studying the behaviour of mixtures of water with apolar solutes.  

The structural arrangements of the solute close to the solvent and
the changes in the structural properties of water due to the solvent
are also related to the behaviour of solubility.
For apolar solute the solubility
is approximately one order of
magnitude smaller than in other solvent like liquid hydrocarbons.
As temperature goes up solubility decreases toward a minimum  
and then sharply increases~\cite{crovetto}. 
At supercritical conditions water becomes a good solvent
for rare gases.
This behaviour has been
interpreted in the
computer simulation studies of 
Guillot and Guissani~\cite{guillot} as
an interplay between 
the energetic term which favours the 
solubility of larger solutes and the
entropic term which depresses the solubility at increasing size.

Only recently it has become possible to perform experimental
studies of the structural correlations in aqueous solutions
of non polar molecules with large neutron and synchrotron 
radiation facilities~\cite{neilson,filipponi1,filipponi2,mar}. 
In particular neutron diffraction studies performed on dilute
mixtures of rare gases in supercritical water~\cite{mar} 
evidenced that the strong increase of the solubility with temperature
could be related to the changes in the hydrogen bond network taking
place in water approaching supercritical conditions. 

A number of computer simulation studies have been performed since the 
pioneering work of Geiger, Rahman and Stillinger~\cite{geiger-rahman} on
aqueous solutions~\cite{guillot,rossky,vaisman,forsman,puhovski, %
skipper,young,lynden-bell,ghosh,paschek}.
Most model potentials introduced to reproduce the properties
of liquid water by computer simulation are not able to predict 
the modification of the hydrogen bond network of water
in the supercritical states evidenced
by experimental results~\cite{soper}. 
If polarizability is taken into account there is a good
improvement in the agreement with the experiments~\cite{jedlo1}. 

Computer simulation of the temperature
dependence of the structural properties of
water and rare gases mixtures from ambient
to supercritical conditions with the use of 
the polarizable potential BSV
have been recently performed~\cite{degrandis}.
This potential was proposed for pure 
water by Ruocco and Sampoli~\cite{r+s}
and later adapted by Brondholt, Sampoli and Vallauri~\cite{bsv} with a more 
refined fit of the potential parameters. The BSV potential is able
to reproduce qualitatively the changes in the site-site correlation
function of water~\cite{jedlo1} in the supercritical region. 
With the BSV potential
at very low fixed concentration~\cite{degrandis} 
a close agreement 
with the neutron diffraction
experiments performed in the same supercritical
conditions~\cite{mar} is found. 

We now consider the effects of increasing both concentration
and size of the solute atoms at ambient and at supercritical conditions
in a computer simulation of aqueous solutions with the BSV potential.
We introduce as a further refinement of the model
the polarizability of the solute atoms which is expected to play a non
negligible role in the case of solutes of large size.
We are motivated in this study by preliminary experiments that have been 
performed in supercritical water
at increasing rare gas concentration~\cite{private}. 
In the structural studies of aqueous solutions
the role of the concentration
of the solute on the microscopic structure of the systems
has been studied for different polar species 
with some controversial interpretation of the 
results~\cite{vaisman,puhovski}. In particular 
the question concerns how the increase of the
solute concentration could improve the ordering
of the solvent.

In the following we present results of MD simulations
performed on solutions of argon and xenon. For the water-argon system
experimental results at supercritical
conditions will become available in a short time~\cite{private}. 
Simulation on 
xenon can help to clarify the effect of changing the size of the solute.

\section{Computer simulation with solute polarizability}

Mixtures of water and rare gases have been simulated by  
Molecular Dynamics (MD) in the microcanonical ensemble. 
The interaction of 
the water molecules is determined by the polarizable
potential BSV~\cite{r+s,bsv}. This potential has the same geometry 
as the TIP4P~\cite{tip4p}
with a polarizable dipole moment placed on the center of mass
of the molecule. 
The parameters are the two positive charges on the hydrogens, the
Lennard-Jones potential parameters on the oxygen and the position
of the negative charge displaced with respect to the oxygen.
The polarizability of water is introduced
by an induced dipole $p_i=\alpha E_i$
calculated from the local electric field $E_i$ with
an iterative procedure and
by assuming an isotropic polarizability fixed to the value for water
molecules~\cite{r+s} $\alpha=1.44$~\AA $^3$. 
In the following we use for water
the parameters determined in ref.~\cite{bsv}.
The parameters of
the Lennard-Jones potential and the polarizability constants $\alpha$
for the rare gases used in the present simulations
are $\epsilon/k_B=125.0$~K, $\sigma=3.415$~\AA, $\alpha=1.586$~\AA$^3$
for $Ar$ and
$\epsilon/k_B=232.0$~K, $\sigma=3.980$~\AA, $\alpha=4.0$~\AA$^3$ for
$Xe$.
The usual Lorentz-Berthelot mixing rules have been employed.

The simulations have been carried out
with the minimum image convention and a cut-off of the interactions
at half of the box length. 
The reaction field has been used
to take into account the long range part of the
electrostatic interactions. 
The thermodynamical points explored range from 
ambient to supercritical conditions. 
In the simulations presented here
a total number of 256 particles
is simulated with
the number of water molecules which varies depending
on the concentration of the solute. 
The volume of the box is adjusted to obtain the chosen density.
In the following we will present results obtained
at ambient conditions and at the supercritical temperature
$T=673$~K and density $\rho=0.331$~g/cm$^3$.
The solute concentrations investigated
are $x=$ $1:40$, $1:30$, $1:20$ and $1:15$. 

\section{Structure of the aqueous solutions}

\subsection{Structure at room temperature}

Our results on the site correlation functions of the
water-argon system indicate that
the increase of the concentration of argon has little effect 
at room temperature on the structure of water. 

The effect appears more relevant
when we consider xenon. The oxygen-oxygen site correlation functions for
both cases are shown in Fig.~1
for the highest studied concentration of solute 1:15 and for pure water.
While with Ar only an increase
of the height and a slight modification of the first peak is observed, 
for increasing fraction of Xe the second and also
the third shells of neighbors are enhanced, as clearly seen
in the figure. In the inset of the figure we report the
coordination
number of the first shell. It changes from $n_{OO}=3.95$ 
at concentration $1:40$ of argon to $n_{OO}=4.45$ at 
concentration $x_{ar}=1:15$. 
The change is more relevant
when we consider xenon 
for which we get $n_{OO}=4.55$      
at $x_{xe}=1:40$ and $n_{OO}=5.21$ for concentration of $x_{xe}=1:15$.

The changes in the oxygen-oxygen first peak height 
could be interpreted from two diiferent poin of view. It
can be related to an enhancement of the water ordering at 
increasing solute concentration but it can also be considered 
as an effect of an enlargement of the local density of water~\cite{vaisman}. 
While the average density of solvent decreases at increasing number
of solute atoms, the fraction of solvent
involved in the coordination shells decreases more slowly than
the average. 
We shall comment below on this by comparing these results
with those obtained in the supercritical region.

The loss of ordering in the water-solute correlation as
the rare gas concentration increases is 
evident from the changes in the arrangement of the solute
atoms around the water molecules. 
The presence of a well defined peak in the solute-solute pair correlation
functions, shown in Fig.~2 for the case of
argon, indicates that
the solute atoms are preferentially close each other. The first peak
increases at increasing concentration
while the first peak of 
the solute-oxygen and the solute-hydrogen pair correlation functions decreases
at increasing concentration as shown in Fig.~3-4.
In  Fig.~2 we also observe for the higher concentration  
a slight shift of the first peak with a shortening of the minimum approach
distance, a more marked shift is observed for the second shell.

At the lowest concentration explored ($x_{ar}=1:40$) the peak 
position of both the
$g_{OAr}(r)$ and the $g_{HAr}(r)$ are located approximately at 
$3.5$~\AA\ according to
the idea that the solute is interstitial in the hydrogen bond network of the
solvent and it is equidistant between oxygens and hydrogens.

By increasing the argon concentration the first peak of 
the $g_{OAr}(r)$ goes down 
indicating that the solute atoms tend to stay more away from water as
there are more of them in the system. 
An estimate of the corresponding first shell coordination numbers
indicates that the number of oxygens close to an argon atom
ranges from $19$ to $16$ upon increasing solute concentration
from $1:40$ to $1:15$.

In the $g_{HAr}(r)$ (Fig.~4) a shoulder is present 
around $4$~\AA\ due to the presence of hydrogens radially oriented with respect
to the solute.
In the $g_{HAr}(r)$ the first peak merges with the shoulder and
at concentration of $1:15$ the first coordination shell of
solute around hydrogens is not well defined anymore. 
These results clearly indicate the tendency to demix of the solutes.
The hydrophobic effect becomes more marked when solute atoms are
added and affects mainly the first shell of the solute
around the water molecules.

With a solute of larger size like xenon the effect 
is even more enhanced at increasing concentration, as shown 
in Fig.~5-6. 
The height of the first peak of $g_{OAr}(r)$ decreases dramatically
with the xenon concentration.
In the case of xenon the estimate
of the first shell coordination numbers
is less precise. We have however obtained a
number of oxygens close to a xenon atom which
ranges from $20$ to $13$ upon increasing solute concentration
from $1:40$ to $1:15$. 
Consistently with what we found for the oxygen-oxygen
coordination number (see inset of Fig.~1) the decrease of nearest neighbours
around the solute is more marked for xenon.
We observe finally that  
the first shell of xenon around
hydrogens is already ill defined at the lowest concentration.

\subsection{Structure in the supercritical region}

We consider now the high temperature region, which
corresponds to the conditions of supercritical water ($T=673$~K and
$\rho=0.331$ g/cm$^3$). 
 
For the lowest argon concentration investigated here
the BSV model shows agreement with experiments performed with
isotopic substitution~\cite{degrandis}. The comparison with
the experimental results is
shown in the bottom panel of Fig.~7.

The change of solute concentration has little effect on the water-solute
structure, at variance with the
ambient conditions where the addition of 
solute atoms increases the tendency of the system to demix.
In Fig.~7-8 we observe only a slight broadening
of the oxygen-solute first peak at increasing concentration.

By comparing Fig.~8 with Fig.~5 we note
that the increase of concentration 
at ambient conditions seems to have an 
effect on the oxygen-xenon structure similar
to the effect of the temperature. However the interpretation
is very different if we look to the water site correlation functions. 
In Fig.~9 
the oxygen-oxygen structure at increasing solute concentration
are compared 
at ambient and supercritical conditions.
This comparison indicates that
at ambient temperature
the increase of the first peak of the $g_{OO}(r)$ 
is at least partially due to an ordering effect 
induced by the solute concentration.
In the supercritical region this effect is completely disappeared
due to weakening of the tetrahedral local order of water. 

In Fig.~10 are reported the $g_{OO}(r)$ at supercritical
conditions for mixtures of Ar and Xe at the concentration 1:15. They
are compared with the pure water. The comparison with Fig.~1
show that the effect of increasing the size
of the solute is much less evident that at ambient conditions. Moreover
the presence of the solute slightly reduces and shifts the peak,
a tendency which is opposite to the one at ambient conditions.

\section{Conclusions}

The use of a polarizable potential for water allowed a realistic
study of 
the structural changes induced by the addition of an apolar solute
to water from ambient up to supercritical 
conditions where non polarizable potentials are known to fail
in reproducing water behaviour. As a further refinement
a polarizable potential for the solute was also introduced in
our simulation.

The increase of the solute concentration 
at ambient conditions causes a strong tendency to demix.
The local order of water seems also to increase. These effects  
are more marked on increasing size of the solute.

It is evident that the tetrahedral network of water plays
an important role in the demixing process since  
the weakening of the
hydrogen bond network which takes place at   
supercritical conditions favours the solubility of the
rare gases in water making almost negligible the effect
of increasing the solute concentration and/or the size of the solutes. 

Comparisons of our results with ongoing experimental studies on similar
systems can help to shed light on the important phenomena
related to the hydrophobic hydration.

\newpage

\section*{Figure captions}

\noindent
Fig.~1. Comparison of the oxygen-oxygen pair correlation 
functions at room temperature of pure water (continuous line) with
aqueous solution of 1:15 Ar (dotted line)
and 1:15 Xe (long dashed line). 
In the inset: coordination number of the first shell of
oxygen-oxygen as function of the concentration $x$
of argon (circles) and xenon (triangles). 

\bigskip

\noindent
Fig.~2. Argon-argon pair correlation functions at room temperature  
at concentration of Ar: $1:40$ (continuous line) and $1:15$ (long dashed line).

\bigskip

\noindent
Fig.~3. Oxygen-argon pair correlation functions at room temperature at 
different concentrations of Ar: 
$1:40$, $1:30$, $1:20$ and $1:15$
from the bottom.

\bigskip

\noindent
Fig.~4. Hydrogen-argon pair correlation functions at room temperature at 
different concentrations of Ar: 
$1:40$, $1:30$, $1:20$ and $1:15$
from the bottom.

\bigskip

\noindent
Fig.~5. Oxygen-xenon pair correlation functions at room temperature at 
different concentration of Xe: 
$x_{xe}=$ $1:40$, $1:30$, $1:20$, $1:15$
from the bottom.

\bigskip

\noindent
Fig.~6. Hydrogen-xenon pair correlation functions at room temperature at 
different concentration of Xe: 
$x_{xe}=$ $1:40$, $1:30$, $1:20$, $1:15$
from the bottom.

\bigskip

\noindent
Fig.~7. Oxygen-argon pair correlation functions at $T=673$~K at 
different concentration of Ar: $1:40$, $1:30$, $1:20$ and $1:15$
from the bottom. The open circles are the experimental results~\cite{mar}.

\bigskip

\noindent
Fig.~8. Oxygen-xenon pair correlation functions at $T=673$~K at 
different concentration of Xe: $1:40$, $1:30$, $1:20$ and $1:15$
from the bottom.

\bigskip

\noindent
Fig.~9. Oxygen-oxygen pair correlation functions 
at $T=300$~K on the left and at 
$T=673$~K on the right with 
different concentration of Xe: $1:40$ (continuous line), $1:15$ 
(long dashed line). For $T=673$~K the two curves are indistinguishable.

\bigskip

\noindent
Fig.~10. Oxygen-oxygen pair correlation functions 
at $T=673$~K  at concentration 1:15 Ar (open triangles)
and Xe (open square) compared with pure water (bold line).
In inset a blow up of the first peak is reported.

\newpage

\begin{figure}[t]
\setlength{\epsfxsize}{100mm}\epsfbox[75 260 380 410]{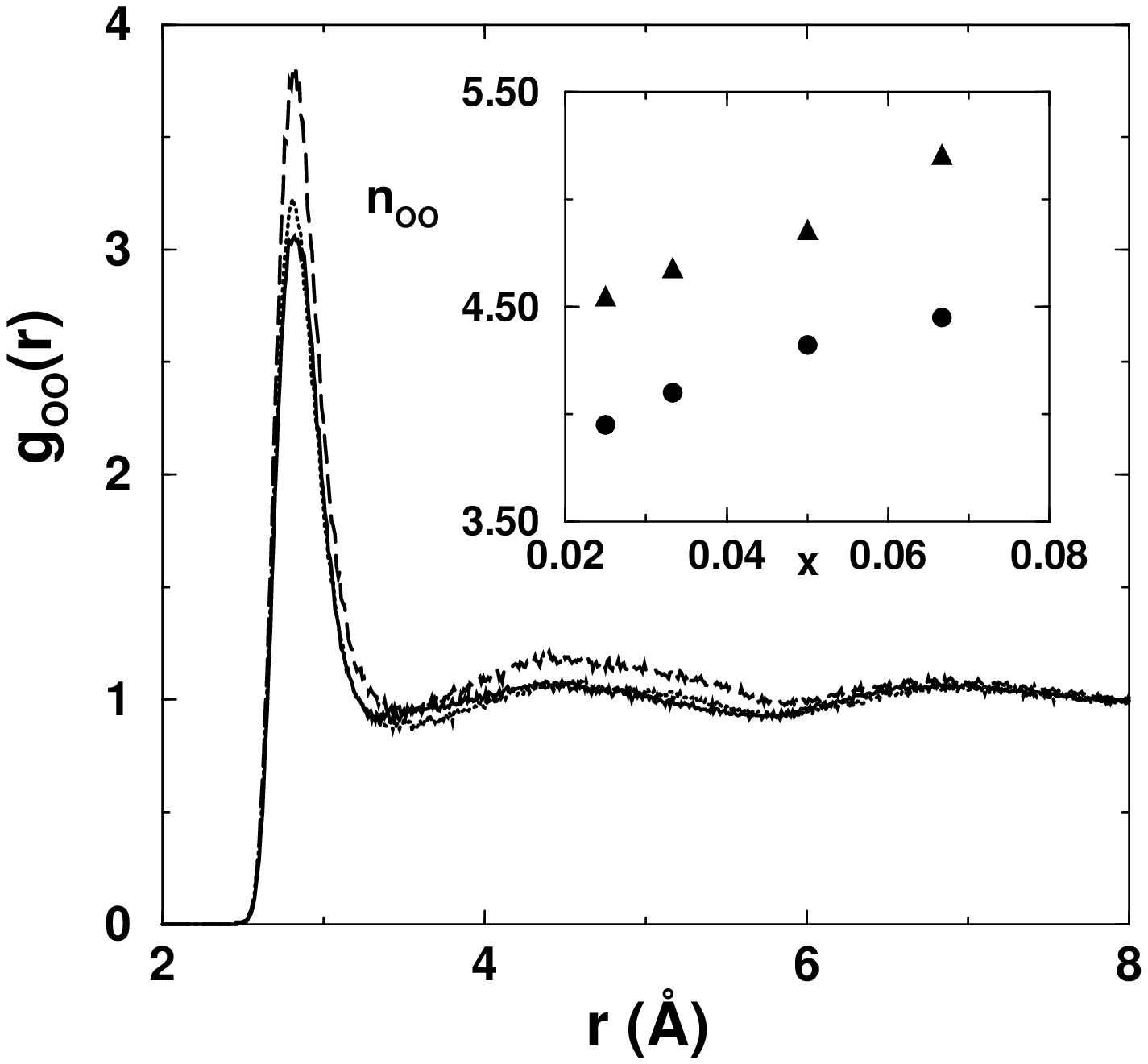}
\vspace{8.truecm}
\protect\label{fig:1}
\end{figure} 
{\bf Figure 1 - P. Cristofori et Al.}
\newpage
\begin{figure}[t]
\setlength{\epsfxsize}{100mm}\epsfbox[75 260 380 410]{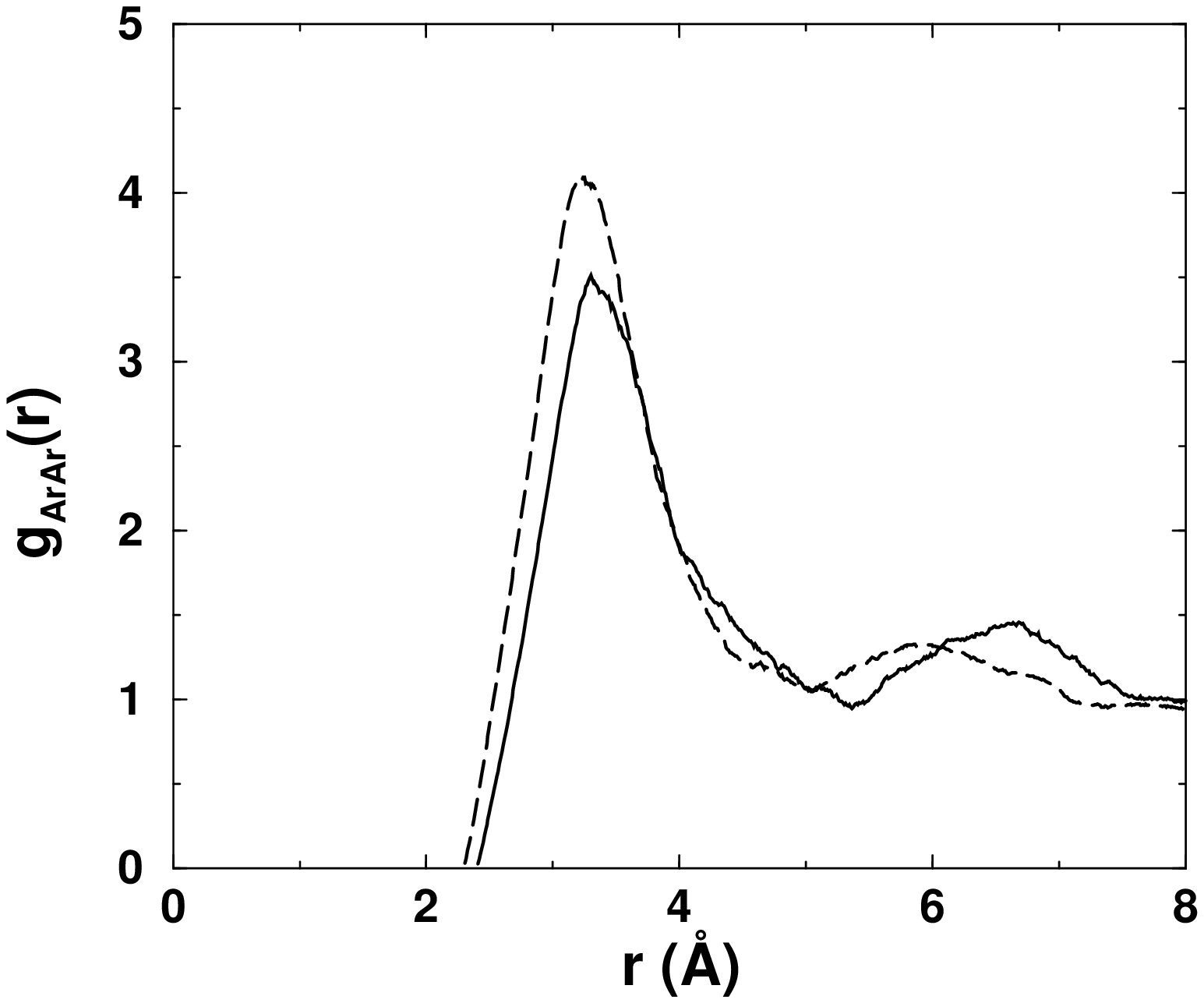}
\vspace{8.truecm}
\end{figure} 
{\bf Figure 2 - P. Cristofori et Al.}
\newpage
\begin{figure}[t]
\setlength{\epsfxsize}{100mm}\epsfbox[75 260 380 410]{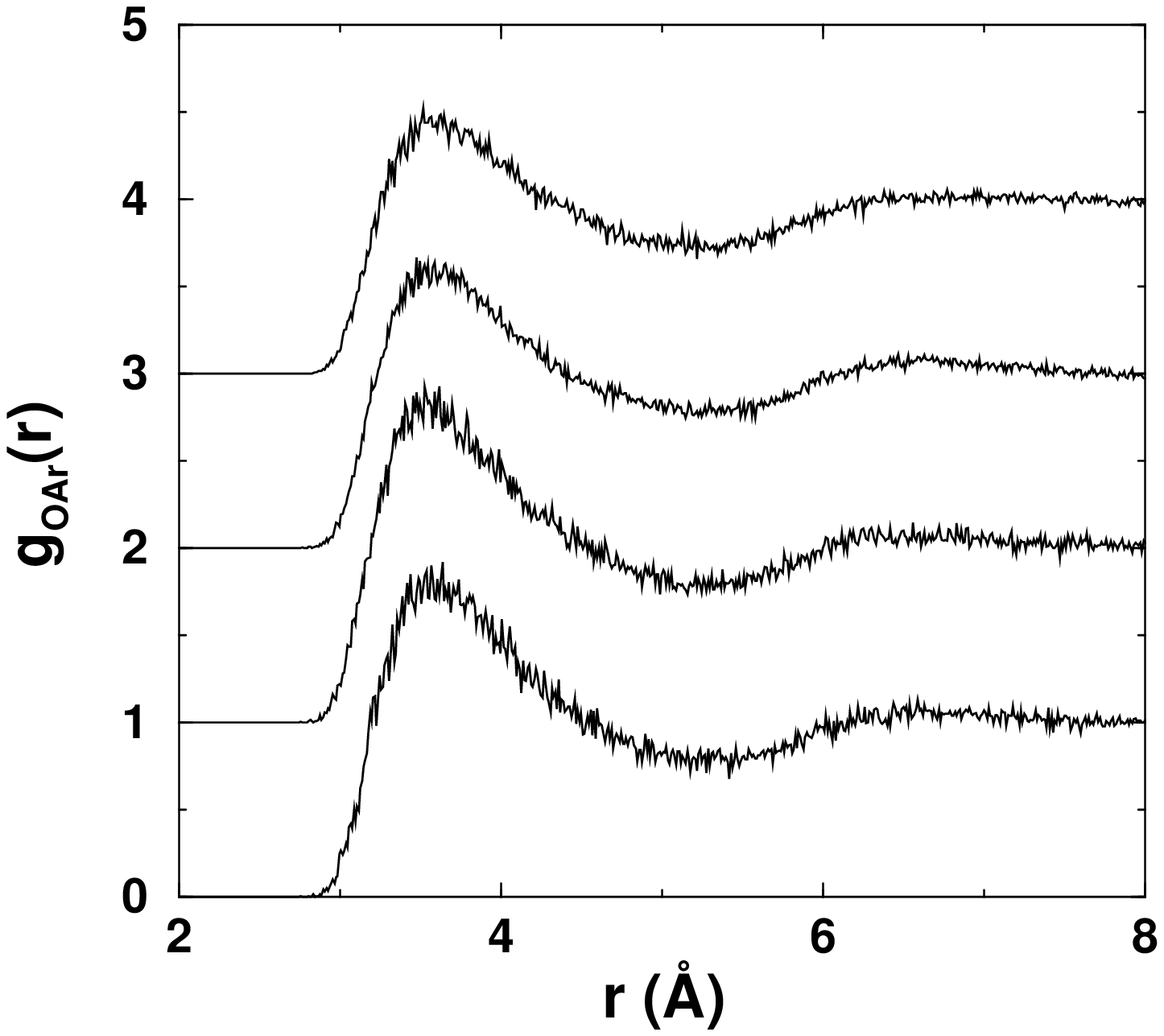}
\vspace{8.truecm}
\end{figure} 
{\bf Figure 3 - P. Cristofori et Al.}
\newpage
\begin{figure}[t]
\setlength{\epsfxsize}{100mm}\epsfbox[75 260 380 410]{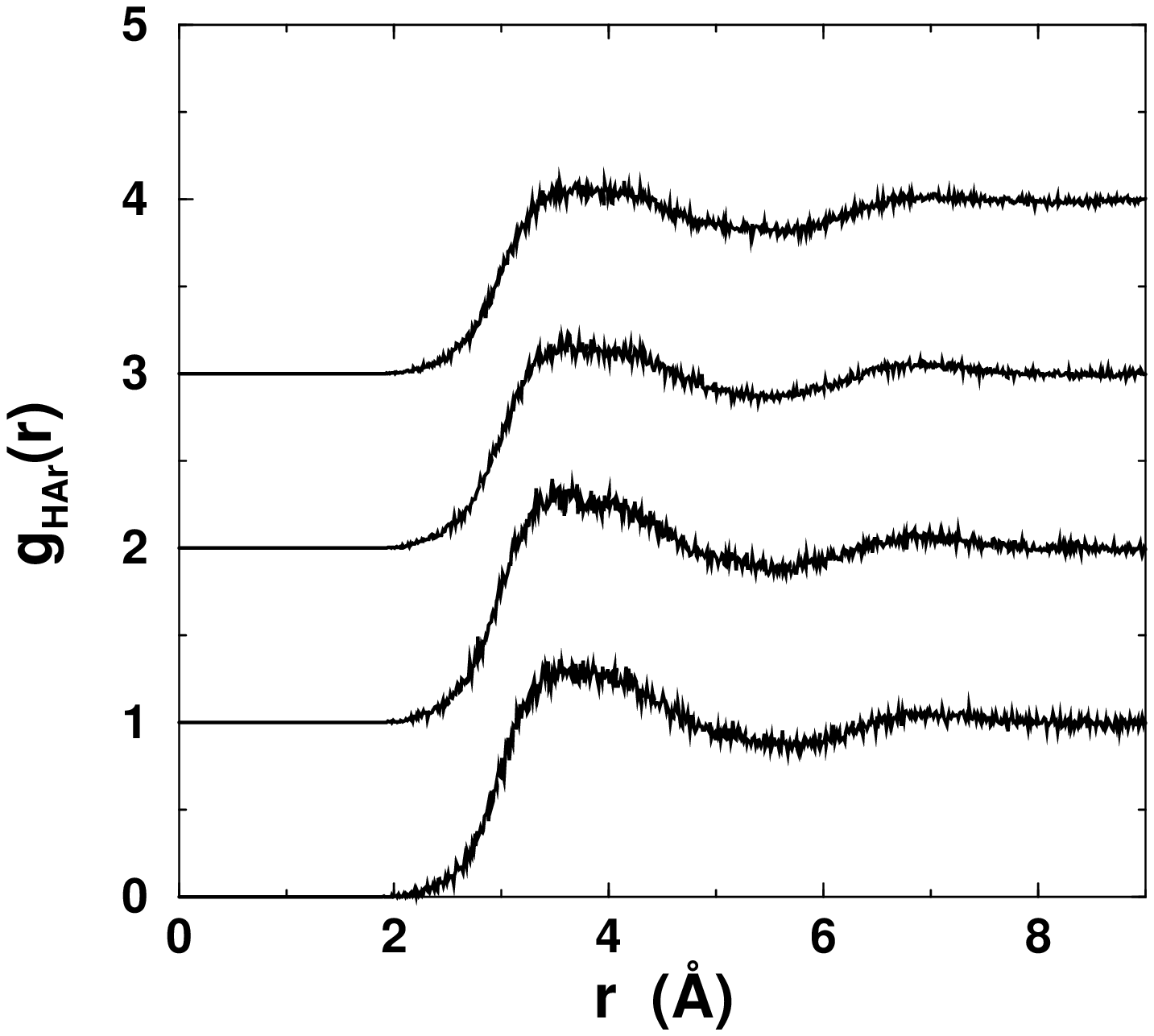}
\vspace{8.truecm}
\end{figure} 
{\bf Figure 4 - P. Cristofori et Al.}
\newpage
\begin{figure}[t]
\setlength{\epsfxsize}{100mm}\epsfbox[75 260 380 410]{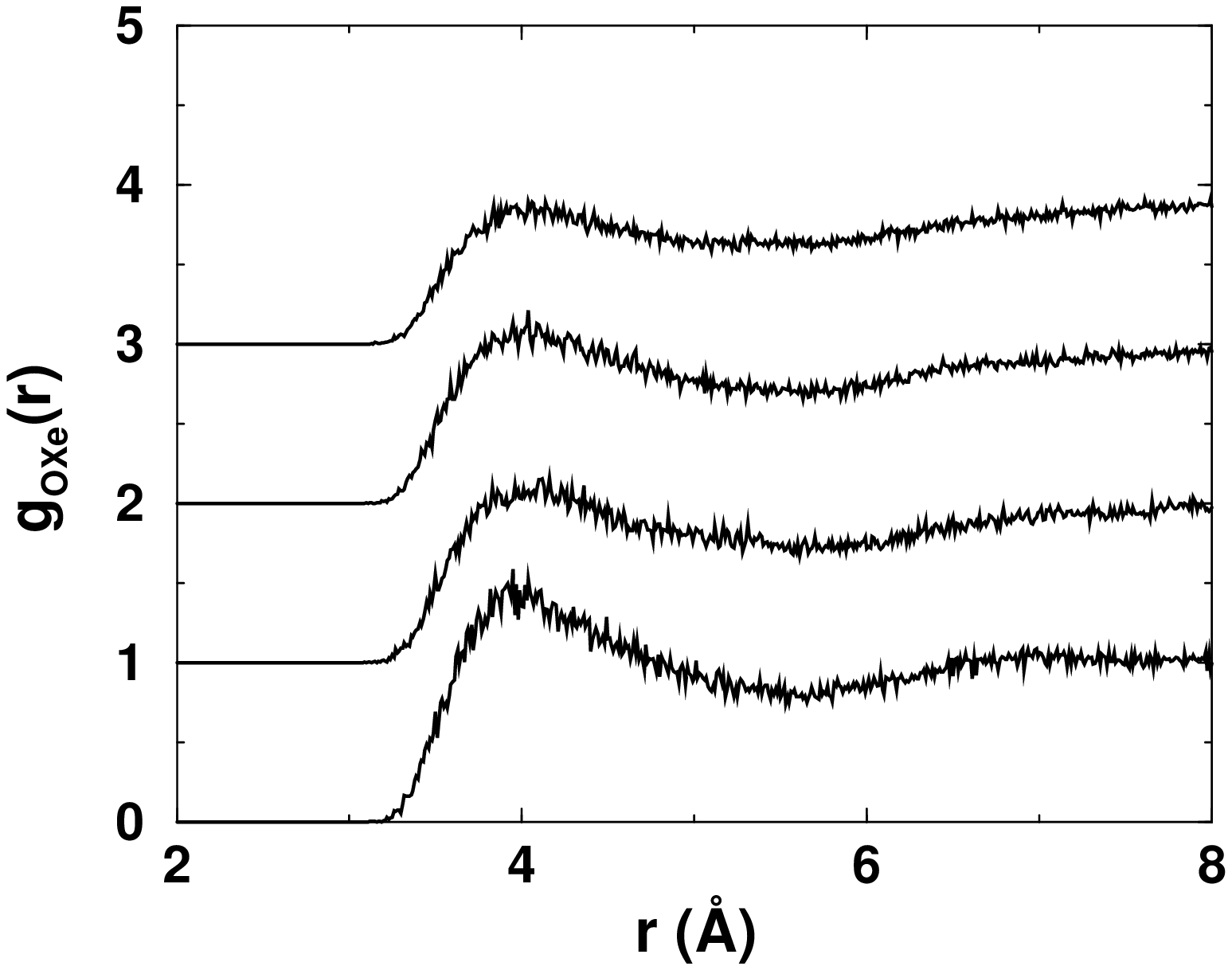}
\vspace{8.truecm}
\end{figure} 
{\bf Figure 5 - P. Cristofori et Al.}
\newpage
\begin{figure}[t]
\setlength{\epsfxsize}{100mm}\epsfbox[75 260 380 410]{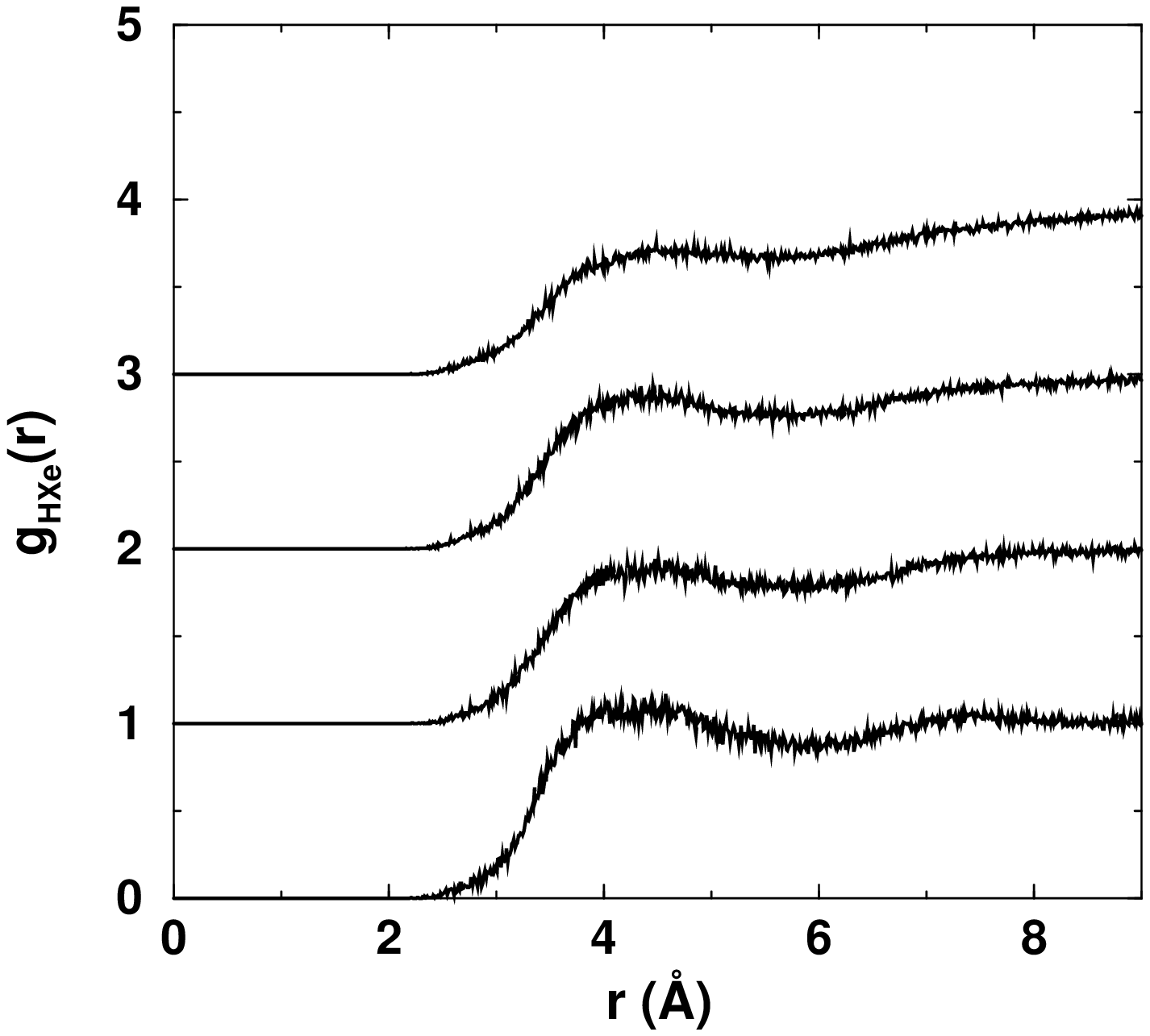}
\vspace{8.truecm}
\end{figure} 
{\bf Figure 6 - P. Cristofori et Al.}
\newpage
\begin{figure}[t]
\setlength{\epsfxsize}{100mm}\epsfbox[75 260 380 410]{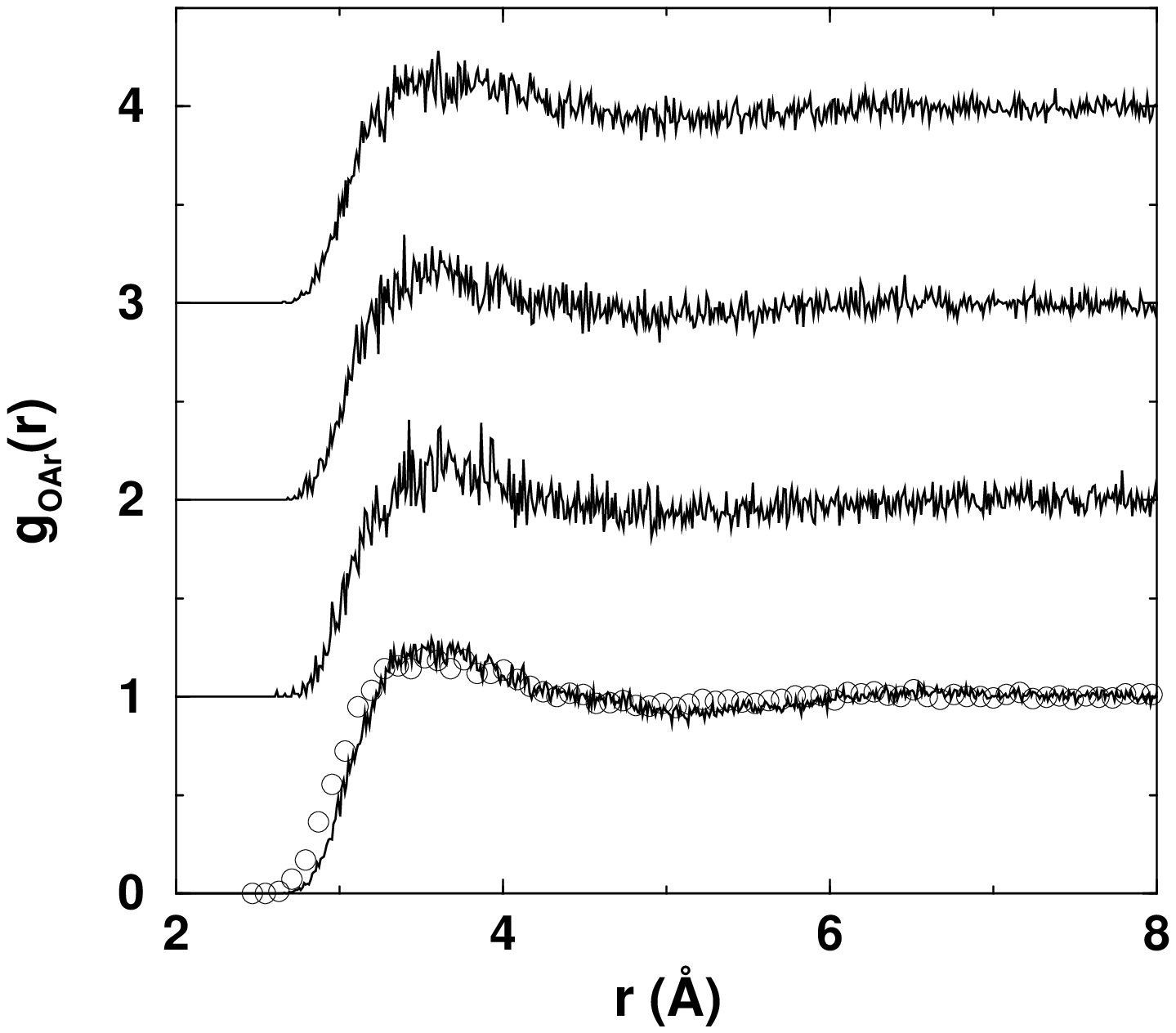}
\vspace{8.truecm}
\end{figure} 
{\bf Figure 7 - P. Cristofori et Al.}
\newpage
\begin{figure}[t]
\setlength{\epsfxsize}{100mm}\epsfbox[75 260 380 410]{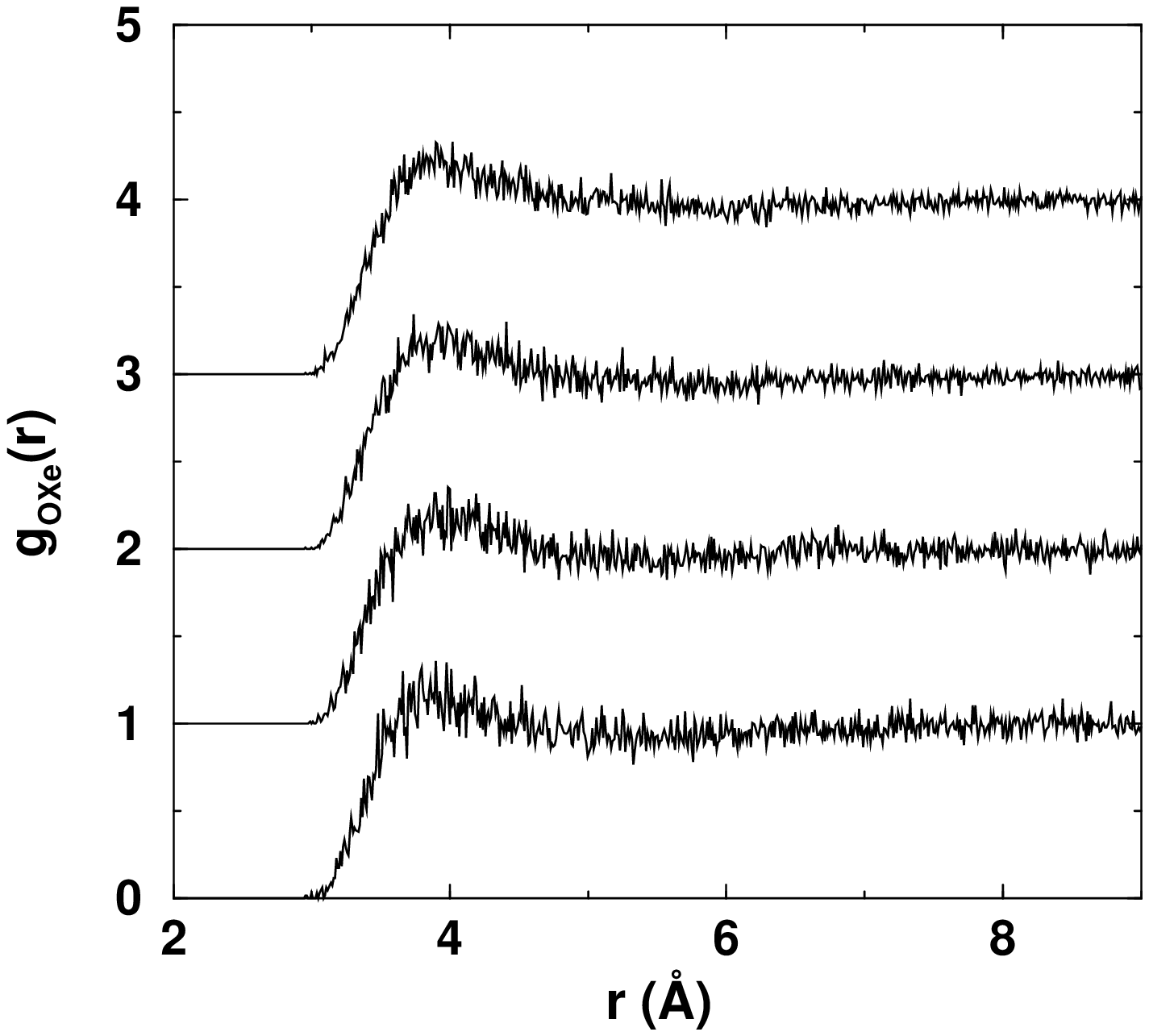}
\vspace{8.truecm}
\end{figure} 
{\bf Figure 8 - P. Cristofori et Al.}
\newpage
\begin{figure}[t]
\setlength{\epsfxsize}{100mm}\epsfbox[75 260 380 410]{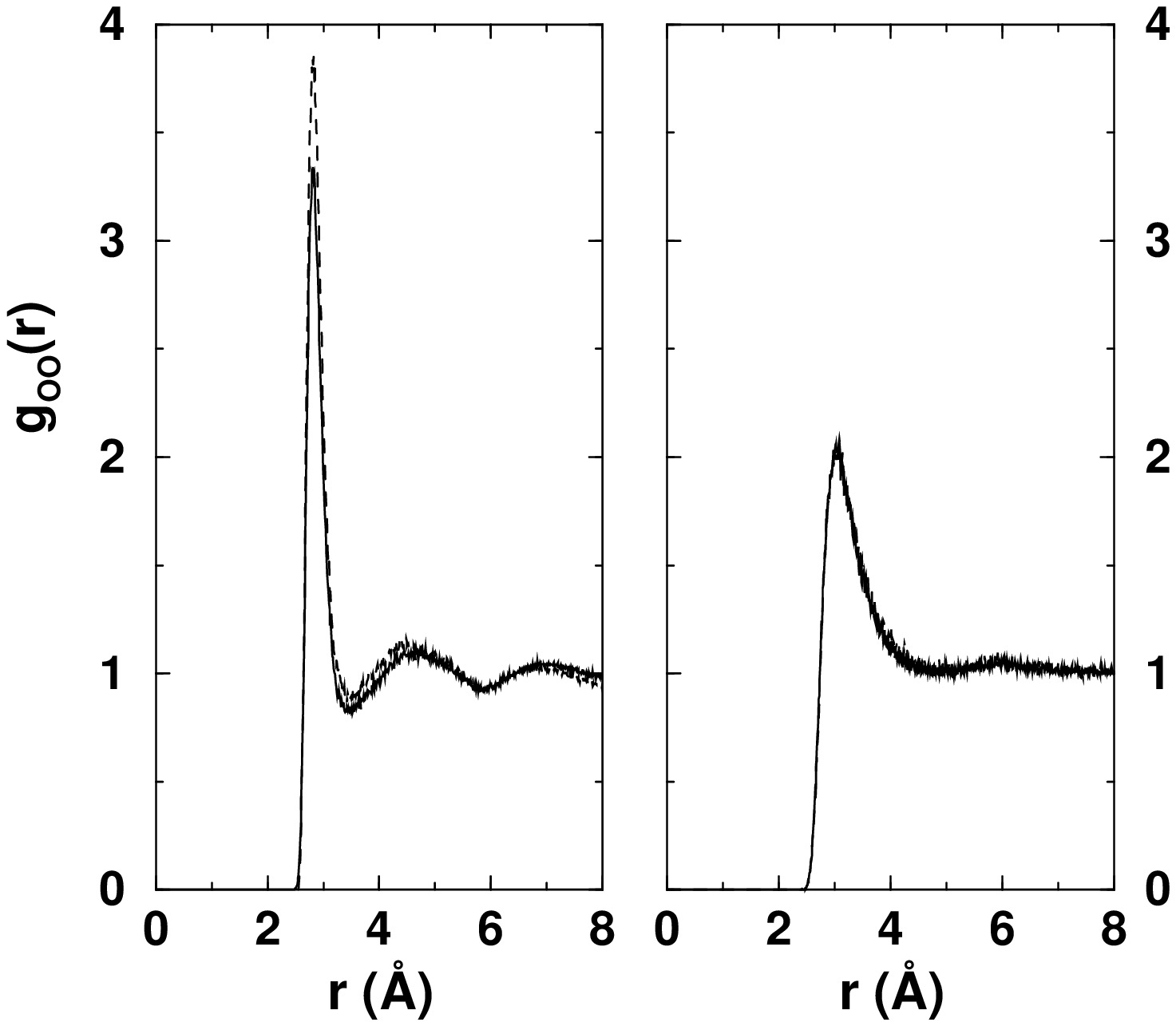}
\vspace{8.truecm}
\end{figure} 
{\bf Figure 9 - P. Cristofori et Al.}
\newpage
\begin{figure}
\setlength{\epsfxsize}{75mm}\epsfbox[75 260 380 410]{fig10.eps}
\vspace{8.truecm}
\end{figure} 
{\bf Figure 10 - P. Cristofori et Al.}

\end{document}